# TDOA assisted RSSD based localization using UWB and directional antennas

Waldemar Gerok[†], Jürgen Peissig
Leibniz Universität Hannover
Hannover, Germany
[†]waldemar.gerok@ikt.uni-hannover.de

Thomas Kaiser
Universität Duisburg-Essen

**Abstract:** This paper studies the use of directional antennas for received signal strength difference (RSSD) based localization using ultra-wideband and demonstrates the achievable accuracy with this localization method applied to UWB. As introduced in our previous work the RSSD localization is assisted with one Time Difference of Arrival (TDOA) estimation. The use of directional receiving antennas and an omni-directional transmitting antenna is assumed. Localization is performed in 2D. Two localization approaches are considered: RSSD using statistical channel model and fingerprinting approach. In the case of statistical channel model simulations are performed using Matlab. In the case of fingerprinting approach localization is done based on real indoor-measurements.

## I. Introduction

Due to its huge bandwidth UWB offers the possibility to achieve very fine time resolution and very precise range estimation can be performed even in an indoor environment with heavy multipath [1-3]. One disadvantage of the time of flight based UWB approaches is the complexity of realization of such systems, resulting from necessity of precise time synchronization, correct detection of first arriving signal component etc. An alternative approach is based on received signal strength (RSS) which offers lower localization accuracy [4, 5] but has also several important advantages. One important feature of RSS is that no precise timing like synchronization between Base Stations (BS) is needed, the RSS values are often available resulting in cheap realization of RSS-based localization and making RSS-based approaches an attractive alternative. To combine the advantages of TOF and RSS hybrid approaches can be found in the literature (e.g. [5, 6]). As shown in the literature adding time measurements to RSS-based positioning can significantly improve the localization accuracy.

In this paper a scenario is considered where a Mobile User (MU) has to be localized using several BS's with known positions. The MU is transmitting an UWB signal, BS's can receive the signal and estimate the RSS, two BS's are estimating TOF [7, 8]. It is further assumed that the transmit power and transmit time are not known exactly. In this case TDOA[1] [1, 9] and RSSD [10, 11] can be calculated and used in the localization procedure. It is additionally assumed that BS's are equipped with directional antennas, the MU is equipped with an omni-directional antenna. Directional antennas are widely used in the localization [2, 12-15]. One of the advantages is the possibility to utilize the directivity in the localization algorithm itself other advantage comes from indirect use of the directivity (e.g. to increase the SNR). In [16] a cognitive interrogator network is proposed, which uses the dynamic orientation of the radiation beams based on the historical MU movements (for more details s. [16]).

---

[1]Sufficient synchronization between the respective BS's is assumed.

Especially in the case of indoor-localization where heavy multipath propagation influences the localization performance utilization of the directional antennas can help to attenuate the reflection components of the signal [14] if the antenna is aligned properly. Similar discussions can be found in [3] where an influence of the antenna on TOF estimation of UWB signals is demonstrated. The aim of this paper is to study the influence of the directional receiving antennas also on the RSSD based indoor localization using UWB (for the transmitter an omni-directional antenna is assumed). The achievable accuracy is of interest.

## II. Statistical Channel Model

In the case of the statistical channel model, in this work simulations are performed using Matlab. First the omni-directional antenna case is explained, in the next step directional antennas on the receiver side are considered and simulations are performed using Matlab. Some results on RSSD based positioning using UWB can be found also in [17]. Using the statistical channel model, the received signal for the omni-directional antenna case can be described as [4]:

$$P_i = P_0 - 10\alpha\log_{10}\left(\frac{d_i}{d_0}\right) + \beta_i \qquad (1)$$

With $P_i$ denoting the received power in dBm at the $i$-th BS, $P_0$ denoting the received power in dBm at the reference distance $d_0$ and $d_i$ denoting the distance between the MU and the $i$-th BS. The number of BS's estimating RSS is $N$. The variable $\beta$ describes the shadow-fading and is assumed as Gaussian distribution with standard deviation $\sigma_\beta$. The variable $\alpha$ is the path-loss exponent. In the case of RSS-base localization distance $d_i$ can be estimated using measured value of $P_i$ and the knowledge of the transmit power. If the transmit power is not known difference of the RSS measured at the BS's can be utilized in the localization [10, 11]. In this case the RSSD does not depend on the transmit power itself [11]:

$$P_{ij} = P_i - P_j = 10\alpha\log_{10}\left(\frac{d_j}{d_i}\right) + \beta_i - \beta_j \qquad (2)$$

Expression (2) can be rewritten as [11]:

$$P_{ij} = P_i - P_j = 5\alpha\log_{10}\left(\frac{(x-x_j)^2 + (y-y_j)^2}{(x-x_i)^2 + (y-y_i)^2}\right) + \beta_i - \beta_j \qquad (3)$$

Here $(x, y)$ describes the position of the MU, which has to be estimated, $(x_i, y_i)$ is the known position of the $i$-th BS. We chose the coordinate system such that both BS's performing TDOA measurement lie on x-axis and the origin of the coordinate system is half the way between these BS's.

As can be seen several measurements between different BS-pairs are needed to estimate the position of the MU. For that purpose a Least Squares (LS) approach can be utilized [11] in which a minimization of a sum $Q(x, y)$ is performed [11]:

$$Q(x,y) = \sum_{i<j}\left(P_{ij} - 5\alpha\log_{10}\left(\frac{(x-x_j)^2 + (y-y_j)^2}{(x-x_i)^2 + (y-y_i)^2}\right)\right)^2 \qquad (4)$$

Since UWB localization is considered the localization error of TOF-based positioning is much lower than the error of RSS-based positioning (cf. [1], [4]) in the case when perfect synchronization and calibration of the system and LOS conditions are assumed. If additionally to RSS-values one TDOA estimation is available the search in (4) can be restricted to the respective hyperbola [7] given by (e.g. [9], [18]):

$$x(y) = (0.5c\Delta t_{kl})\sqrt{1 + y^2/(s_{kl}^2 - (0.5c\Delta t_{kl})^2)} \tag{5}$$

Here $\Delta t_{kl}$ denotes the TDOA estimation, $2s_{kl}$ denotes the distance between the respective $k$-th and $l$-th BS's involved in the TDOA estimation and $c$ is the speed of light. Using (5) the expression (4) can be rewritten as (cf. [7]):

$$\hat{y} = \arg\min_{y} \sum_{i<j} \left( P_{ij} - 5\alpha \log_{10}\left( \frac{\left((0.5c\Delta t_{kl})\sqrt{1+y^2/(s_{kl}^2-(0.5c\Delta t_{kl})^2)} - x_j\right)^2 + (y-y_j)^2}{\left((0.5c\Delta t_{kl})\sqrt{1+y^2/(s_{kl}^2-(0.5c\Delta t_{kl})^2)} - x_i\right)^2 + (y-y_i)^2} \right) \right)^2 \tag{6}$$

Ones the $y$-position is estimated $x$-position can also be found using (5).

### A. Propagation Constants

For the signal strength based localization two parameters are especially important: shadow fading and path-loss exponent (cf. [5-7]). Those propagation constants depend also on the utilized antennas, as can be seen from [19] antenna directivity has an influence on the shadow fading and on the pass-loss exponent of an UWB signal. Authors in [19] studied different antenna combinations: both transmitting and receiving antenna are omni-directional (Omni/Omni), both antennas are directional (Dir/Dir) and the case where one of the antenna is directional and other omni-directional (Omni/Dir) case. Comparison of the propagation constants show that shadow fading is reduced and path-loss exponent is increased in Omni/Dir case compared to Omni/Omni case. Reducing the shadow fading and increasing the path-loss exponent results in improved accuracy for signal strength based localizations. In [7] we suggested to use a similar scheme as in [16] and to orientate the directional antennas toward the MU also for RSS/RSSD based positioning. The orientation can be based on historical position estimations of the MU. In our study, when new position estimation is performed we dynamically update the orientation of the receiving antennas and align the main lobe of the antennas toward the new position estimation in the case if statistical channel model is considered. Misorientation of the directional antennas will remain small if the position estimation result has only small error and the antenna movement between two position estimations is small enough in this case shadow fading and path-loss exponent can contribute to the localization accuracy for RSSD case .

### B. Directional Antenna

As mentioned before, for the BS's directional antennas are assumed. For the MU an omni-directional antenna is considered. To only demonstrate the effect of the antennas on the localization performance the dependency of the antenna radiation pattern on the frequency range of interest is assumed to be small enough and is neglected, the radiation pattern is approximated similar to [15] as:

$$g(\varphi) = G \cos \varphi \tag{7}$$

The gain is chosen as $G = 6.5$. To include the effect of the antenna on the "measured RSSD" following expression is used (cf. [15]):

$$P_{ij}^{dir} = g(\varphi_i) - g(\varphi_j) + 10\alpha \log_{10}\left(\frac{d_j}{d_i}\right) + \beta_i - \beta_j \qquad (8)$$

Here $\varphi_i$ is denoting the azimuth angle between the real MU position and $i$-th BS orientation (for BS perfectly orientated toward the MU $\varphi = 0$).

### C. Simulation Results

To study the influence of the directional antenna on the TDOA assisted RSSD localization we assume an area of 8 m x 8 m and a total number of $N = 8$ BS's performing only RSS estimation, two additional BS's perform only TDOA estimation. The movement of the MU is chosen according to model used in [20]. The MU is moved in a square area of 7 m x 7 m in which also a position search is performed. To visualize the localization error a relatively short track is generated, the total length is ca. 18 m, for this track the "pause time" before choosing a new random position to move to is set to "0". It is further assumed that the initial position of the MU is known and the orientation of directional receiving antennas is correct for the first position of the MU (calculation of the positioning error is made excluding the first position for both Omi/Dir and Omni/Omni case to allow fair comparison of the results). Simulations are performed using Matlab in which the value of the path-loss exponent and the standard deviation of shadow fading constant are chosen according to [19]. The standard deviation of TDOA estimation error is exemplary chosen as 330 ps. The localization is done 2 times per second unless other noted.

It is assumed that the misorientation of the directional antennas is small enough and the propagation constants correspond to the Omni/Dir case. As mentioned before if one TDOA estimation is available the LS search can be simplified to (cf. [7]):

$$\hat{y} = \arg\min_{y} \sum_{i<j} \left( P_{ij}^{dir} - 5\alpha \log_{10}\left(\frac{\left((0.5c\Delta t_{kl})\sqrt{1 + y^2/(s_{kl}^2 - (0.5c\Delta t_{kl})^2)} - x_j\right)^2 + (y - y_j)^2}{\left((0.5c\Delta t_{kl})\sqrt{1 + y^2/(s_{kl}^2 - (0.5c\Delta t_{kl})^2)} - x_i\right)^2 + (y - y_i)^2}\right) - g(\varphi(x(y), y)_i) + g(\varphi(x(y), y)_j) \right)^2 \qquad (9)$$

In the first step the $y$-position of the MU can be estimated in the next step (5) can be used also for the $x$-position estimation. In (9) the azimuth angle $\varphi(x(y), y)$ between the assumed/tested MU position and the orientation of the respective BS antenna must be included in the calculation of the MU position (if e.g. the $i$-th BS antenna is pointing to the tested MU position $\varphi(x(y), y)_i = 0°$).

The BS orientation itself is chosen according to MU position estimation from previous localization step (s. Fig. 1).

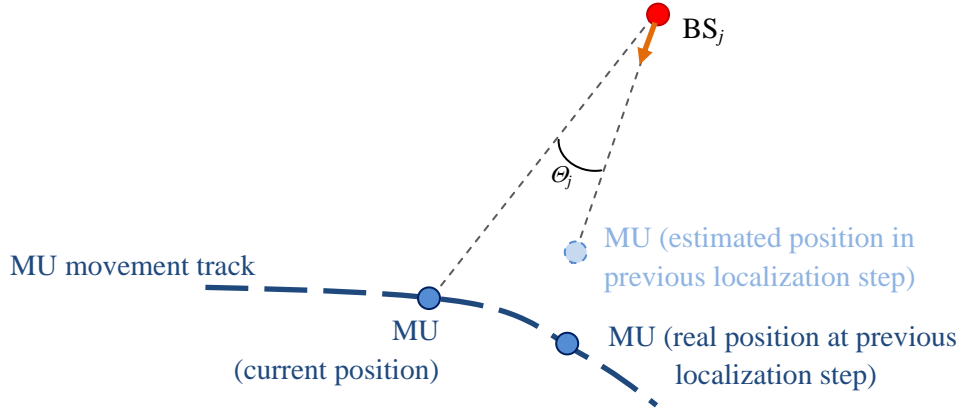

Fig. 1: The misorientation angle $\Theta_j$ is exemplary shown for $BS_j$. The orientation of directional antennas is dynamically updated.

If no TDOA estimation is available expression (4) can be extended with the respective values $P_{ij}^{dir}$ and with consideration of the antenna gains for the azimuth angle $\varphi(x,y)_i$ and $\varphi(x,y)_j$:

$$Q(x,y) = \sum_{i<j} \left( P_{ij}^{dir} - 5\alpha \log_{10}\left(\frac{(x-x_j)^2 + (y-y_j)^2}{(x-x_i)^2 + (y-y_i)^2}\right) - g(\varphi(x,y)_i) + g(\varphi(x,y)_j) \right)^2 \quad (10)$$

In [7] we demonstrated the influence of a single TDOA estimation on the RSSD-based positioning. In this paper the simulations are done using also directional receiving antennas. Fig. 2 shows the localization results for TDOA assisted RSSD based localization.

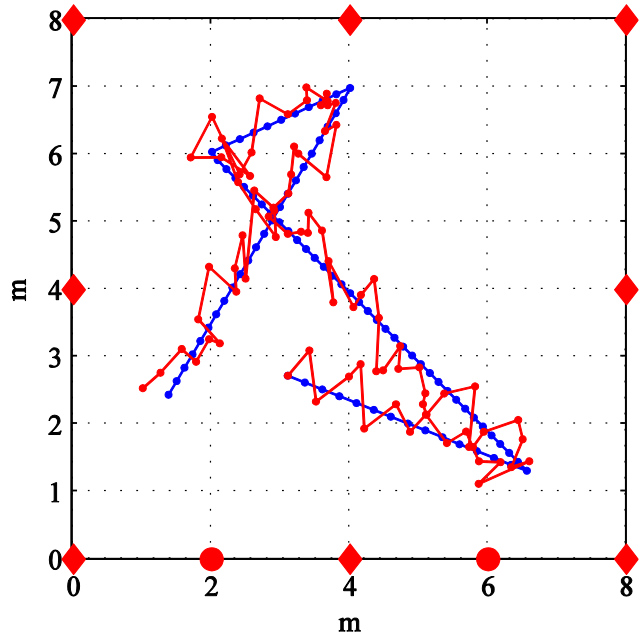

Fig. 2: Simulation result of TDOA assisted RSSD localization. Directional receiving antennas are assumed. Real track is shown in blue line, estimated position is shown in red line. BS's estimating only RSS shown in red diamonds, BS's used for only TDOA estimation shown in red circles (assumption, that the first position of the MU is known). Value of the path-loss exponent and the standard deviation of shadow fading constant are chosen according to [19] Omni/Dir case.

The Root Mean Squared Error (RMSE) of TDOA assisted RSSD localization for exemplary chosen track using directional antennas is 0.33 m. The value[2] of the standard deviation of $\Theta$ calculated over all positions and all antennas is only 4.1°. If the simulations are repeated assuming omni-directional receiving antennas RMSE of the localization increases to 0.59 m, the values of the path loss exponent and the standard deviation of shadow fading constant are chosen according to [19] Omni/Omni case.

If no TDOA estimation is available and only LS RSSD-based localization is performed the localization error for the directional receiving antenna case increases to 0.46 m, the value of the standard deviation of $\Theta$ calculated over all positions and all antennas is 5.5°. Fig. 3 shows the respective simulation results. If the simulations are repeated for the omni-directional receiving antenna case RMSE of the localization is 0.95 m.

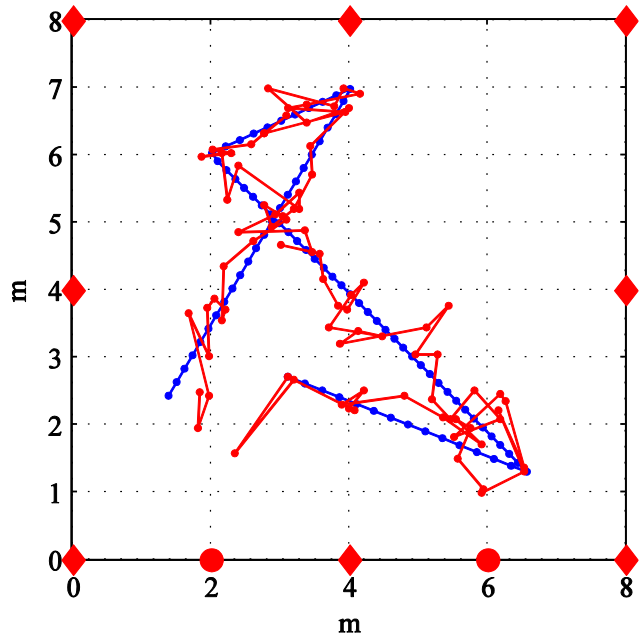

Fig. 3: Simulation result of RSSD localization. Directional receiving antennas are assumed. Real track is shown in blue line, estimated position is shown in red line. BS's estimating only RSS shown in red diamonds, BS's used for only TDOA estimation (shown in red circles) are not utilized in the position estimation (assumption, that the first position of the MU is known). Value of the standard deviation of shadow fading constant and path-loss exponent are chosen according to [19] Omni/Dir case.

An alternative to the utilization of the statistical channel model fingerprinting based localization can be applied [21, 22].

## III. Fingerprinting

In the case of the fingerprinting we perform real indoor measurements using an experimental UWB Test Bed in this paper. The paper extends also our work [8] to a directional antenna case. In the following the orientation of the directional antennas will be fixed.

Some work on RSS-based UWB localization can be found in [21].

---

[2] If the simulation is repeated and the localization is done only 1 time per second the respective value of standard deviation of $\Theta$ calculated over all positions and all antennas increases to 5.1°.

In the RSS-based fingerprinting approach two phases are performed [22]:

1. In the offline phase the reference values $P_{j\_ref}(x, y)$ of RSS at the BS's are collected for known reference locations of the MU $(x, y)$.
2. In the online phase the localization is performed and the position of the MU is estimated. To perform the localization the LS approach can be chosen, in which the reference values $P_{j\_ref}(x, y)$ are compared to the measured values $P_{j\_meas}$ of RSS and the respective Euclidean distance is calculated for the reference positions of the MU [22]:

$$E(x,y) = \sqrt{\sum_{j}^{N} \left(P_{j\_mess} - P_{j\_ref}(x,y)\right)^2} \quad (11)$$

In the simple implementation the reference position, which minimizes the Euclidian distance, can be chosen as the estimated position of the MU.

In [10] RSSD based fingerprinting is performed for WLAN. In the case of RSSD the values $P_{ij}$ are used instead of $P_i$ in (11). We utilize all $P_{ij}$ with $1 \leq i < j$ in the following calculations [10, 11].

In the following the position of the MU is calculated in two steps: first we calculate the position of the MU using only RSSD and the LS fingerprinting approach (coarse estimation), in the next step the position of the MU is refined by projecting the estimated coarse position on the hyperbola corresponding to the measured TDOA (for more details s. [7, 8]). The resulting position is chosen as final position estimation of MU.

### A. Localization Set Up

The localization set up consists of a Digital Phosphor Oscilloscope (DPO) with the input channels connected to Vivaldi-antennas, the DPO and Vivaldi-antennas are emulating four BS's. The MU is emulated using an Arbitrary Waveform Generator (AWG) an Amplifier and a monoconical antenna. The system is controlled via a PC. The gain of the Vivaldi antennas is about 5 dB at 3 GHz. The experimental UWB signal is a 128 chip bi-phase modulated pulse train with pulse repetition frequency of 3 MHz, the 10 dB bandwidth of the signal is about 2.3 GHz $< f <$ 3.9 GHz. The peak power of the transmitted pulse is 28 dBm. No active inband interferer is in the direct vicinity of the system. The BS's are positioned at $BS_1$: (0, 0) m; $BS_2$: (3, 0) m; $BS_3$: (0, 3) m; $BS_4$: (3, 3) m. Two metallic objects are added to the set-up in order to increase the reflections on the $BS_4$ and obstruct $BS_4$ for part of the measurements. Fig. 4 shows the set-up.

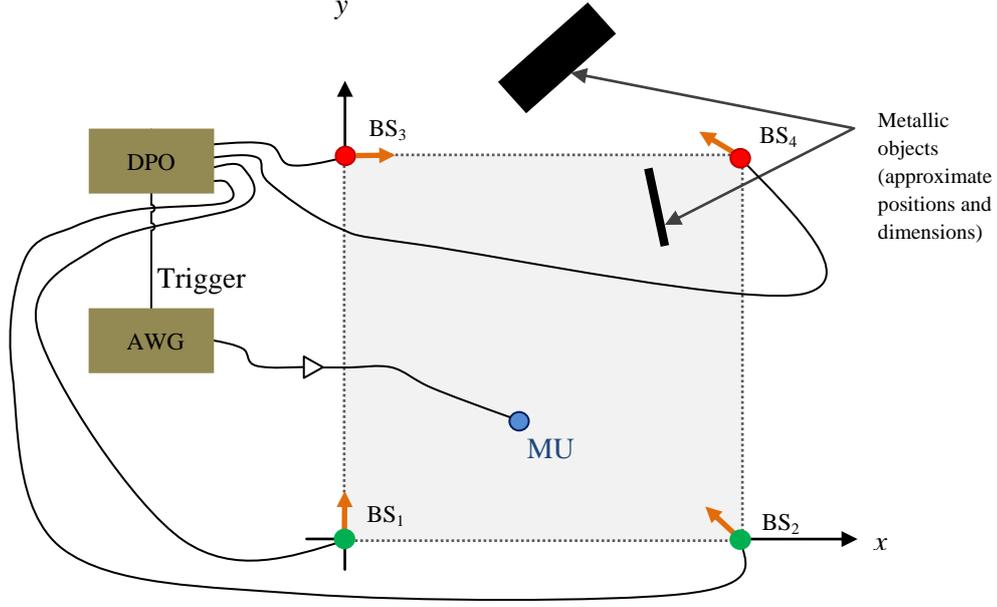

Fig. 4: Localization scenario in the fingerprinting approach. MU is shown in blue, BS's used for only RSS measurements are shown in red, BS's used for RSS and TDOA measurements are shown in green. Orientation of receiving antennas of BS's is shown using orange arrows – the orientation of the antennas is fixed. The area of interest is shown in light gray.

The RSS values are measured on all four BS's, the TDOA is calculated using $BS_1$ and $BS_2$. To calculate the TDOA cross-correlation receiver is implemented in Matlab [1, 2], the respective signals are BP filtered, correlated with a known template and upsampled. The TDOA is estimated as difference of the positions of the peaks of the respective correlation functions. Fig. 5 shows the implementation of the receiver in Matlab.

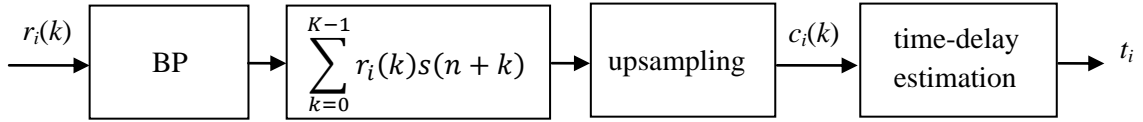

Fig. 5: Implementation of the cross-correlation receiver in Matlab (cf. [1, 2]) for the *i*-th BS.

In the fig. 5 $r_i(k)$ is denoting the signal received at the $BS_i$, $s_i(k)$ is the known template of the transmit signal, $c_i(k)$ denotes the respective cross-correlation result and $t_i$ is the time-delay.

The RSS is calculated using the output of the cross-correlation receiver (similar to [4]):

$$P_i = \frac{1}{t_b - t_a} \int_{t_a}^{t_b} \left(c_i(t)\right)^2 dt \quad (12)$$

Here $t_a$ and $t_b$ are the relevant parts of the received signal. The integration period is chosen as 70 ns.

### B. Localization Results (Measurements)

In the offline phase the RSS values are collected on the four antennas for the predefined positions of the MU. Positions are defined by a square grid of 0.25 m x 0.25 m in the considered 3 m x 3 m area of interest, positions in the direct vicinity to the receiving antennas and the two metallic objects are excluded from the considerations [8]. In the online phase the MU is moved along a circular track with

radius of 1 m. The total number of positions is $M = 48$. The predefined positions of the circular track are:

$$\begin{aligned} x(m) &= 1.5 + \cos(-90° + 7.5°(m-1)) \\ y(m) &= 1.5 + \sin(-90° + 7.5°(m-1)) \end{aligned}$$ (13)

Here $m$ denotes the number of the position on the predefined track. Fig. 6 shows the result of the localization of the MU on the predefined track.

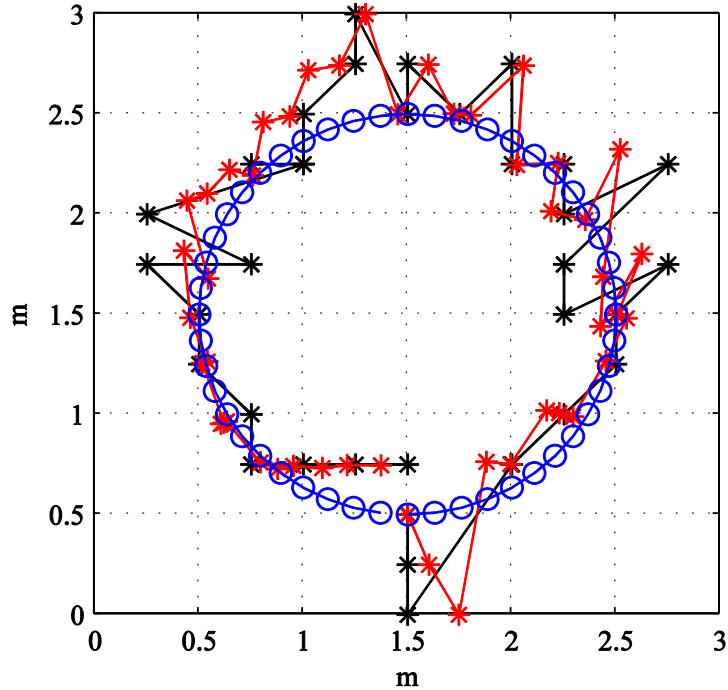

Fig. 6: Localization result of the fingerprinting approach. Real track is shown in blue, TDOA assisted RSSD localization result is shown in red, pure RSSD based localization is shown in black.

The RMSE in the case that TDOA is considered is 21.4 cm, for pure RSSD-based localization the RMSE is 24.9 cm.

## IV. Conclusion

In this work the influence of the directional antennas on RSSD based indoor localization with UWB is demonstrated. As in our previous work additionally to RSSD single TDOA is utilized to assist the localization. Simulations show very promising results in the case of statistical channel model, it was demonstrated that dynamical orientation of the directional antennas has the potential to increase the localization accuracy also for the RSSD-localization case.